\documentclass[a4paper,twocolumn,pre,floatfix,showpacs,aps,10pt]{revtex4-1}
\usepackage{times}
\usepackage{balance}
\usepackage{soul}

\usepackage{graphicx} 
\usepackage{lastpage}

\usepackage{amsfonts,amsmath,amssymb,marvosym}
\usepackage{textcomp}
\usepackage{bm}
\usepackage{graphicx}
\usepackage[latin1]{inputenc}
\usepackage{rotating}
\usepackage{color}
\usepackage{hyperref}
\usepackage{upgreek}

\newcommand{\mrbub}{\mathrm{bub}}
\newcommand{\mrpit}{\mathrm{pit}}
\newcommand{\mrair}{\mathrm{air}}

\newcommand{\mro}{\mathrm{o}}
\newcommand{\mratm}{\mathrm{0}}
\newcommand{\mrl}{\mathrm{l}}
\newcommand{\mrs}{\mathrm{s}}
\newcommand{\mryl}{\mathrm{YL}}
\newcommand{\mrhyd}{\mathrm{hyd}}

\newcommand{\mrtot}{\mathrm{tot}}
\newcommand{\mrint}{\mathrm{int}}
\newcommand{\mrv}{\mathrm{v}}
\newcommand{\mrH}{\mathrm{H}}

\newcommand{\etal}{\textit{et~al.}}

\newcommand{\SIC}{^\circ\!\textrm{C}}

\newcommand{\SIkg}{\textrm{kg}}

\newcommand{\SIm}{\textrm{m}}

\newcommand{\SIcm}{\textrm{cm}}
\newcommand{\SImum}{\textrm{\textmu{}m}}

\newcommand{\beq}[1]{\begin{equation} \eqlab{#1}}
\newcommand{\eeq}{\end{equation}}
\newcommand{\bsub}{\begin{subequations}}
\newcommand{\esub}{\end{subequations}}
\def\bal#1\eal{\begin{align}#1\end{align}}
\def\bsubal#1\esubal{\bsub \begin{align}#1\end{align} \esub}

\newcommand{\eqlab}[1]{\label{eq:#1}}
\renewcommand{\eqref}[1]{Eq.~(\ref{eq:#1})}

\newcommand{\figref}[1]{Fig.~\ref{fig:#1}}

\newcommand{\figlab}[1]{\label{fig:#1}}

\newcommand{\seclab}[1]{\label{sec:#1}}

\begin{document}

\title{Growth control of sessile microbubbles in PDMS devices }

\author{
Andreas Volk,\textit{$^{a}$}
Massimiliano Rossi,\footnote{Corresponding author}$^{\ast}$\textit{$^{a}$}
Christian J. K\"ahler,\textit{$^{a}$}
Sascha Hilgenfeldt,\textit{$^{b}$}
and
Alvaro Marin\textit{$^{a}$} 
}
\thanks{Corresponding Authors: alvarogum@gmail.com, m.rossi@unibw.de}
\affiliation{
{$^{a}$~\textit{Institute for Fluid Mechanics and Aerodynamics, Bundeswehr University Munich, 85577 Neubiberg, Germany, E-mail: a.marin@unibw.de}}\\
{$^{b}$~\textit{Department of Mechanical Science and Engineering, University of Illinois at Urbana Champaign, Urbana-Champaign, USA}
}}

\begin{abstract}
In a microfluidic environment, the presence of bubbles is often detrimental to the functionality of the device, leading to clogging or cavitation, but microbubbles can also be an indispensable asset in other applications such as microstreaming. In either case, it is crucial to understand and control the growth or shrinkage of these bodies of air, in particular in common soft-lithography devices based on polydimethylsiloxane (PDMS), which is highly permeable to gases. In this work, we study the gas transport into and out of a bubble positioned in a microfluidic device, taking into account the direct gas exchange through PDMS as well as the transport of gas through the liquid in the device. Hydrostatic pressure regulation allows for the quantitative control of growth, shrinkage, or the attainment of a stable equilibrium bubble size. We find that the vapor pressure of the liquid plays an important role for the balance of gas transport, accounting for variability in experimental conditions and suggesting additional means of bubble size control in applications.
\end{abstract}
\maketitle

\section{Introduction}
\seclab{introduction}
Bubbles are ubiquitous in our everyday life, \cite{ligerbelair2005} in arts, \cite{prosperetti2004} in engineering, \cite{kantarci2005} and also in microfluidics. Uncontrolled growing bubbles in small devices can fatally disrupt many processes in microfluidics: for example, Shin et al. \cite{shin2003} performed Polymerase Chain Reactions (PCR) in a microfluidic device and found that the main reason for failure was the formation and growth of bubbles, which they managed to control by coating the surface of their device with substances of low permeability to avoid bubble formation.\cite{liu2007} However, in other systems such as carbonated drinks or beer, growing bubbles are desired and understanding their formation and growth is crucial to optimize foam formation,  taste and  texture. \cite{ligerbelair2005, enriquez2013, rodriguez2014} In other applications, the presence of bubbles is necessary, but their growth (or dissolution) must be under control. That is crucial in micro fuel cells, in which $\mathrm{CO_2}$ bubbles are usually generated and need to be dissolved in the surrounding water solution; the presence of other gas species in the liquid can be detrimental to this process, stabilizing the bubbles and preventing their total dissolution \cite{shim2014,Pablo:CO2bubble}.

\begin{figure}[h!]
  \centering
  \includegraphics[width=\columnwidth]{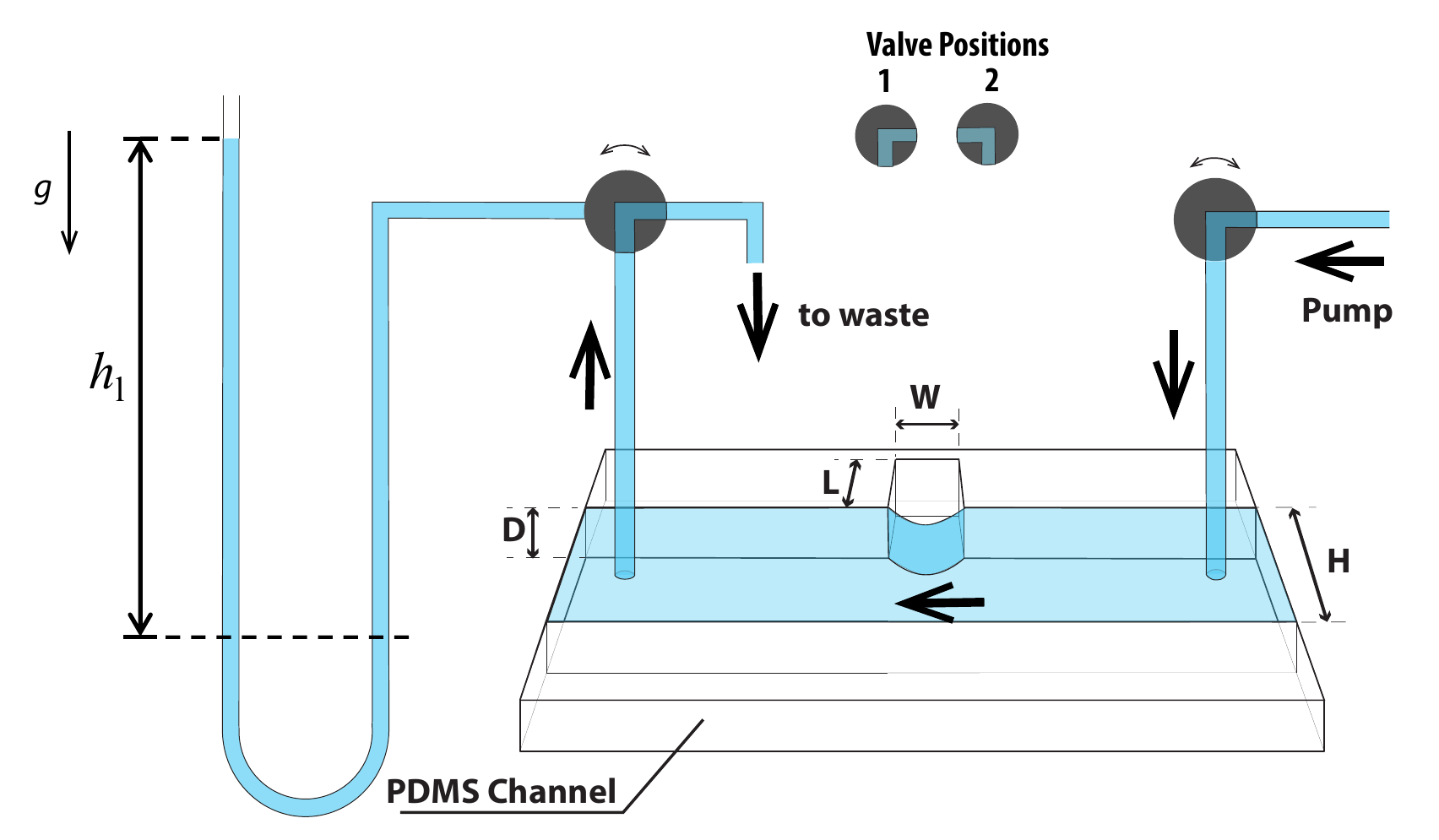}
\caption{\figlab{fig1} Experiments are performed on a PDMS channel filled by a syringe pump (or by a pressurized system). With the two-way valves in position {\bf 1}, the PDMS-microchannel is filled with liquid from the pump. As the liquid enters the channel, an air pocket is trapped in the blind side channel forming a sessile bubble. By switching the valves to position {\bf 2}, the channel is open to the atmosphere through the open tube at the left side of the sketch. The liquid's pressure is adjusted by moving the flexible tube's end to the desired relative liquid level $h_{\mathrm{l}}$.}
\end{figure}

Other applications require bubbles of specific size at known positions. This is the case in acoustic streaming driven by oscillating microbubbles \cite{marmottant2003,marmottant2004,wang2011}: A sessile bubble is stabilized in a blind channel or micro-pit and actuated through a piezo transducer to induce bubble oscillations that generate a steady streaming flow in the liquid. Control of the bubble size and shape is important in order to retain constant flow conditions and a constant bubble response to the chosen driving frequency. On silicon substrates, the bubble is controlled by simply oversaturating the liquid solution \cite{marmottant2006, epstein1950}. The bubble then grows due to oversaturation until the pressure differences are equilibrated by surface tension, as mandated by the Young-Laplace equation. If the system is closed and there are no changes in pressure, concentration or temperature, the bubble should remain stable at a constant radius. Unfortunately that is not the case when working in systems based on polydimethylsiloxane (PDMS): PDMS is by far the most commonly used material for microfluidic applications due to its biocompatibility, ease of fabrication and low price \cite{xia1998}. But it is also a porous medium permeable to gas (and to some degree to liquid) \cite{merkel2000}. The porosity of PDMS has been used as an advantage for promoting fluid motion due to the evaporation through the pore network \cite{randall2005} or to supply oxygen to cells in PDMS-based bioreactors \cite{deJong2006}. However, controlling the growth of bubbles becomes a more complex task. To our knowledge, no strategies have been proposed explicitly in the literature for the control of bubbles in PDMS-based microfluidic devices.

In this paper, we study the stability of sessile microbubbles trapped in blind channels of PDMS devices under different conditions and propose methods based on the control of the hydrostatic pressure inside the device which are easy to implement in any system. Our conclusions are particularly useful for microfluidic applications, which require control of the gas transfer both through a gas-liquid interface as well as a gas-solid interface. 

\section{Experimental Setup}
\seclab{expsetup}

Experiments are performed on a PDMS microchannel with a blind side channel and a tubing system to control the filling and liquid pressure, as sketched in \figref{fig1}. The PDMS device is fabricated using standard soft lithography, as described by Wang \etal ~\cite{wang2012}. The main channel has a rectangular cross-section with a depth {D$\,=100~\SImum$ and a height H$\,=250~\SImum$. The blind side channel, perpendicular to the main channel and sharing the depth D, has a width W$\,=50~\SImum$ and a height L$\,=150~\SImum$. \textcolor{black}{Both main and blind side channel are completely embedded in PDMS. The channel structure is bounded by a PDMS layer of 2 mm thickness towards the glass slide to which the device is attached, and by PDMS layers of $\sim$4\,mm thickness on all other sides, towards the surrounding atmosphere.}
The setup includes two two-way valves and a manometer-style water column to adjust the relative water level $h_\mrl$ and thus the hydrostatic pressure in the system. With the valves in position {\bf 1}, the microchannel is filled with liquid from the pump, remaining unconnected to the water column. Once the channel is filled, an air pocket remains trapped in the side channel forming a quasi-hemicylindrical bubble. In order to study the bubble growth without a net flow across the channel, the valves are then switched to position {\bf 2}, which closes the inlet and connects the outlet to the water column. The liquid's pressure is then adjusted by the relative liquid level $h_{\mrl}$. \textcolor{black}{In our experiments this was done by connecting a flexible tube to the outlet valve (tube connected to left valve in \figref{fig1}). The tube was filled with water up to the open end. Using a vertical positioning system with millimetric resolution, the end of the tube was moved up or down in order to set the relative liquid level to the desired value}. 

\begin{figure}[h!]
  \centering
  \includegraphics[width=\columnwidth]{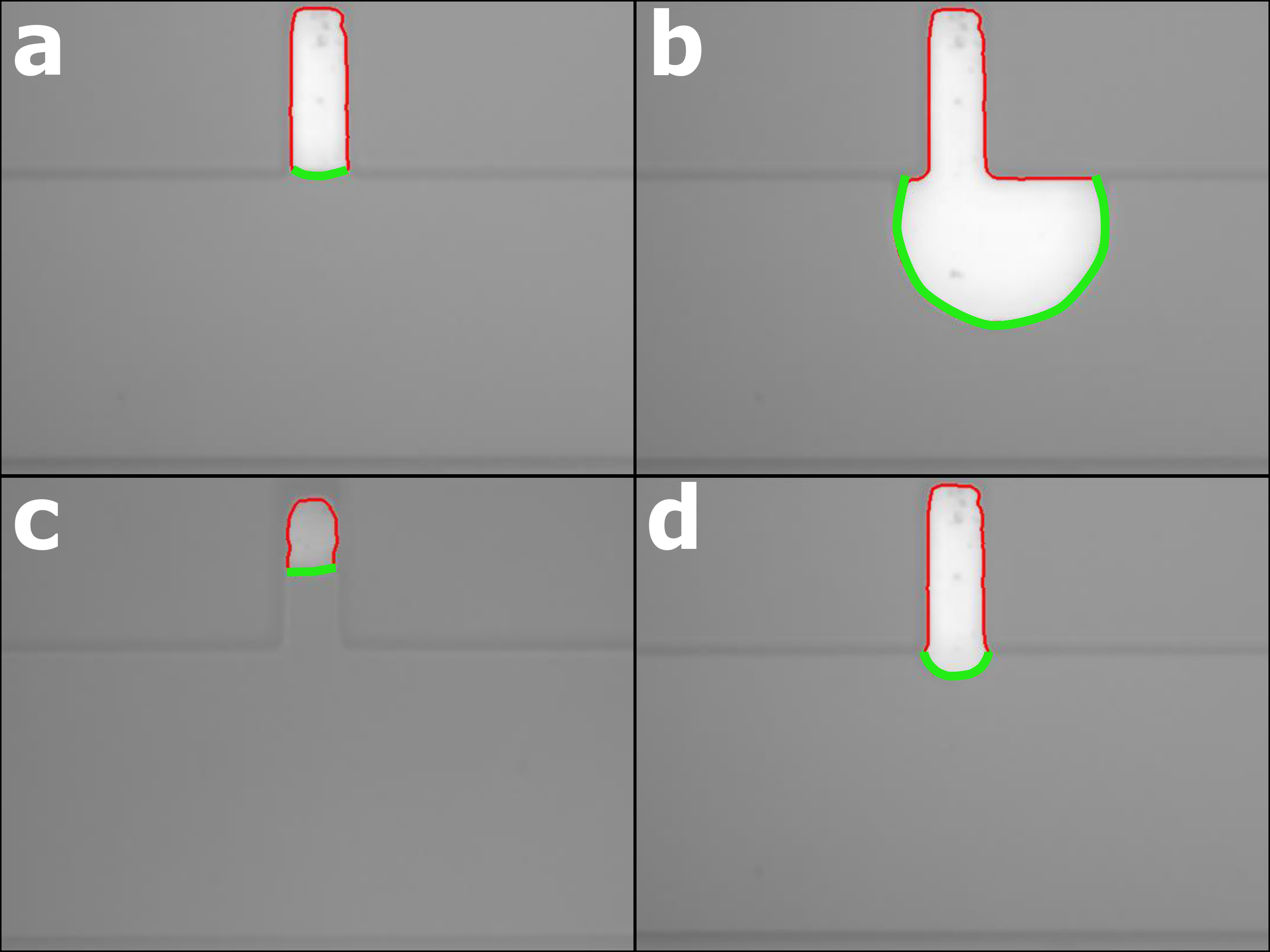}
\caption{\figlab{bubimg} Bubble volume is calculated by fitting a circle to the gas-liquid interface (green line) and by determining the boundary between gas and PDMS (red line). Each experiment starts at $t=0$ with the bubble volume roughly matching the pit's (a). By adjusting the liquid hydrostatic pressure we can obtain growing (b), shrinking (c), or steady bubbles (d).}
\end{figure}

\section{Results}
\seclab{results}

\begin{figure*}[t]
\includegraphics[width=2\columnwidth]{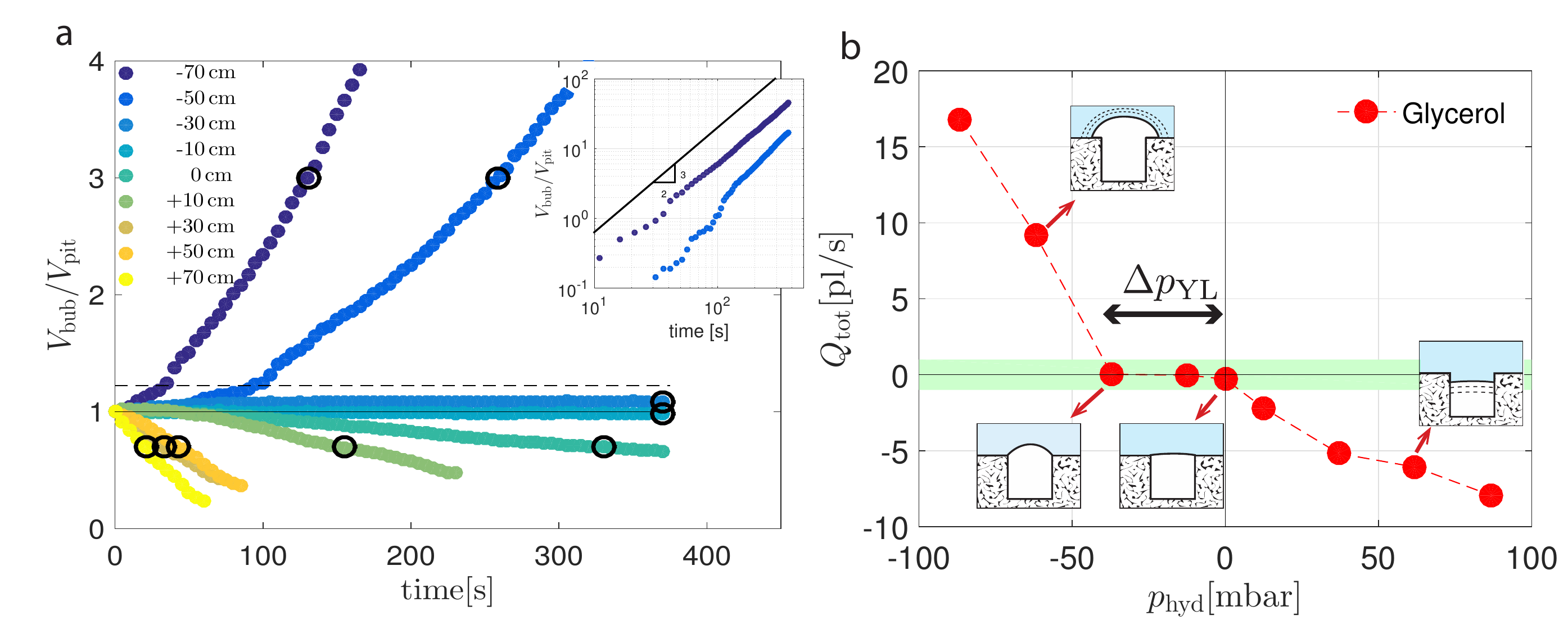}
\caption{\figlab{TVgly} a) Normalized bubble volume as a function of time for different liquid levels $h_l$ in the experiments with glycerol. The liquid in the microchannel is connected to a movable liquid column at different heights to obtain different hydrostatic pressures. The black dashed line corresponds to the volume of a semi-cylindrical bubble with $a=W/2$. The black circles correspond to those points where the slope is taken to define the growth rate $Q_\mrtot$. In the insert a log-log graph shows the $V\sim t^{3/2}$ scaling that growing bubbles follow at large time scales. b) Bubble growth rate $Q_\mrtot$ as a function of the hydrostatic pressure $p_\mrhyd$ in the liquid. The bubble is growing in the upper part ($Q_{\mrtot}>0$) and shrinking in the lower half of the plot ($Q_{\mrtot}<0$). Bubble stability is achieved for the data points within the green area, where \textcolor{black}{$-\Delta p_{YL} < p_{hyd} < 0; \Delta p_{YL} = p_{YL}(a=W/2) - p_{YL}(a=\infty)$ and} the growth or shrinkage is negligible within the experimental time scale.  }
\end{figure*}

Assuming stationary or quasi-stationary conditions, the gas pressure inside the microbubble $p_{\mrbub}$ can be calculated using the Young-Laplace equation:   

\textsc{}
\textcolor{black}{
\beq{BubblePressure}
	p_{\mrbub}= p_{\mratm} +\rho_{\mrl} g h_{\mrl} + C\gamma_\mrl
\eeq
where $p_{\mratm} $ is the atmospheric pressure, $g$ is the gravitational acceleration, $\rho_{\mrl}$ is the liquid density, $\gamma_{\mrl}$ is the liquid's surface tension and $C$ is the bubble curvature. Since the equilibrium contact angle of water and glycerol in PDMS is close to $90^\circ$ \cite{deGennes}, we can assume that the bubble is practically cylindrical and therefore  $C=1/a$, where $a$ is the bubble projected radius.
From now on we will refer to the last term of \eqref{BubblePressure} as \textcolor{black}{$p_{\mathrm{YL}}={\gamma_\mrl}/{a}$} or as Young-Laplace pressure. Considering the hydrostatic pressure $p_{\mrhyd}=\rho_{\mrl} g h_{\mrl}$ the difference between bubble pressure and atmospheric pressure is given by $\Delta p=p_{\mrhyd} + p_{\mryl}$.} Each experiment was performed at constant hydrostatic pressure (constant $h_{\mathrm{l}}$) and started from a microbubble with an almost flat liquid-gas interface (see \figref{bubimg}a). By manipulating the control variables, 
the bubble can be made to grow (\figref{bubimg}b), shrink (\figref{bubimg}c) or reach a steady state in which the bubble interface adopts an intermediate shape between flat and hemicylindrical (\figref{bubimg}d). 

A microscope (Zeiss Axio Imager Z2) and a camera (Zeiss AxioCam HRm) are used to visualize the sessile bubble during the experiments. Each experiment ran for a maximum of 400 seconds, acquiring images at 0.2 frames per second. \textcolor{black}{ For each image, the boundaries of the bubble were extracted with an image processing software and used } to determine the bubble volume $V_{\mrbub}$, the bubble radius $a$, the liquid-gas area $A_\mrl$, and the PDMS-gas area $A_\mrs$ as a function of time. The depth dimension was always assumed to be the nominal depth D of the channel. The bubble radius $a$ was measured performing a circle fit \cite{thomas1989} to the liquid-gas interface (green lines in \figref{bubimg}). Three different liquids have been used: glycerol, air-saturated water, and degassed water. The liquids are chosen to analyze the effects of air solubility and vapor pressure. We used triply deionized water (Milli-Q millipore) for the experiments. Air-saturated water was typically stored in a fridge overnight at circa $10 \SIC$. Water was degassed at 0.07 bar for 30 min with a magnetic stirrer; the sample was then immediately inserted in the PDMS device.

\subsection{Experiments using Glycerol}
\seclab{measurements_glycerol}

The first set of experiments was performed using glycerol as working fluid. Figure \ref{fig:TVgly}a shows the bubble volume $V_{\mrbub}$ as a function of time, for different hydrostatic pressures (or relative liquid levels $h_{\mrl}$). The values of $V_{\mrbub}$ are normalized to the volume of the blind side channel (or pit) $V_{\mrpit}=\mathrm{DWL}$, so that $V_{\mrbub}/V_{\mrpit}=1$ at the beginning of each experiment. Bubbles grow for liquid pressures much lower than atmospheric (negative values of $h_\mrl$). As we increase the hydrostatic pressure the growth decreases until stable bubbles are achieved. In our experiments we found a range of hydrostatic pressures at which the bubble remains stable within our observation time-scale, spanning from 
$-3710~\mathrm{Pa}$ to $-1240~\mathrm{Pa}$ 
(corresponding to $h_{\mrl}=-30\,\SIcm$ to $h_{\mrl}=-10\,\SIcm$ using the glycerol density of $\rho_\mrl=1261\,\SIkg/\SIm^{3}$).



From the curves shown in \figref{TVgly}a, we quantify representative bubble growth rates $Q_\mrtot$ for different hydrostatic pressures. For growing bubbles, the growth rate of each data set is computed at $V_\mrbub/V_\mrpit=3$, averaging over the growth in a time interval of $20$\,s; this is a convenient definition, and the results do not vary significantly when the time interval of the average ensemble is changed. For stable bubbles, the growth rate is zero by definition. For shrinking bubbles, it is quantified at the point where their volume shrinks to 70\% of $V_{\mrpit}$. The respective locations for evaluating $Q_\mrtot$ are indicated as black circles in Fig. \ref{fig:TVgly}a; the values are plotted against $p_\mrhyd$ in Fig. \ref{fig:TVgly}b. 


Bubble stability manifests in Fig. \ref{fig:TVgly}b as a plateau where $Q_\mrtot \approx 0$. In this region, the bubble's curvature decreases as the hydrostatic pressure is increased, so that the bubble's air pressure remains constant and $\Delta{p}^\mrair \approx 0$. Stable bubbles have a hemicylindrical cross section at the smallest $p_\mrhyd$ and an almost flat shape at $p_\mrhyd=0$. Therefore, the plateau's width is well approximated by the value of the Young-Laplace pressure of a $a=\mathrm{W}/2$ semi-cylindrical bubble, $2 \gamma_\mathrm{l}/\mathrm{W}=2520~\mathrm{Pa}$. Note that at $p_\mrhyd=0$ ($h_\mrl=0$) the bubble shrinks slowly since the bubble is not completely flat and the air partial pressure difference $\Delta{p}$ is positive and equal to the Young-Laplace pressure. Beyond this point, both  hydrostatic and Young-Laplace pressure have the same sign and they contribute to increase the bubble's air pressure. Air is then forced through the system and the bubble shrinks into the pit ($Q_\mrtot<0$), eventually vanishing completely after sufficient time. 
%

\subsection{Experiments using Water}
\seclab{measurements_water}


Since most applications use aqueous solutions, two more sets of experiments were performed using air-saturated and degassed water as working liquids. In this case, neither gas diffusion nor evaporation can be neglected. Since the diffusivity of vapor in air is four decades larger than that of air in water ($D_{\mathrm{air}}^{\mathrm{w}}=2\cdot10^{-9}~\SIm^2/s$ and $D_{\mathrm{vapor}}^{\mathrm{air}}=0.282\cdot10^{-4}~\SIm^2/s$), we can safely assume that evaporation is much faster than air diffusion and the bubble is always saturated with vapor. This is indeed the most important difference from the case of glycerol, given that the vapor pressure of glycerol under atmospheric conditions is negligible for our experiments. Therefore, in experiments with volatile liquids, the bubble pressure depends also on the liquid's vapor pressure.

\begin{figure}[t]
  \includegraphics[width=\columnwidth]{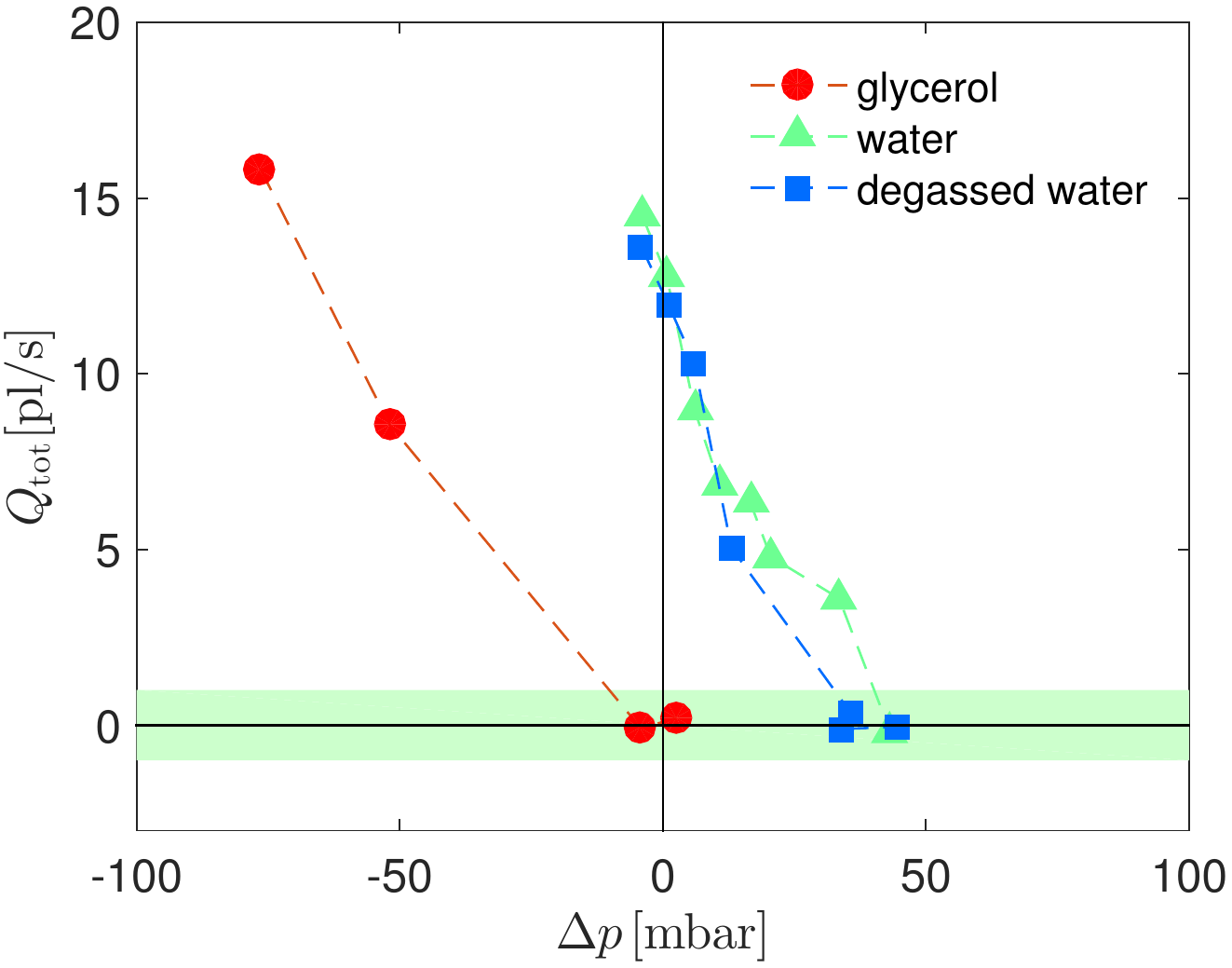}
\caption{\figlab{PQtot} Bubble growth rate $Q_{\mrtot}$ for glycerol, air-saturated water and degassed water as a function \textcolor{black}{$\Delta p$}.} 
\end{figure}
The results are plotted in \figref{PQtot} as a function of \textcolor{black}{$\Delta{p}=p_\mrhyd + p_\mryl$} for $Q_\mrtot > 0$ only. Note that $\Delta{p}$ is constant for stable bubbles (see \eqref{BubblePressure}) so that the plateau present in \figref{TVgly}b is now absent in this representation. In \figref{PQtot} we can see that all experiments have a similar linear dependence on the hydrostatic pressure, differing only in their horizontal position in the plot -- the experiments with water show growth rates that are displaced to higher pressures, independent of the liquid's gas content. We will show below that such a displacement can be explained by taking into account the liquid vapor pressure.

\section{Discussion}
\seclab{discussion}

The most relevant difference between experiments with water and glycerol is the significantly higher, and temperature dependent vapor pressure of the former in comparison to the latter. Consequently, the gas inside the bubble (as well as the air surrounding the set-up) contains a certain amount of vapor, and the air partial pressure difference can be written as 

\begin{equation}
\Delta{p}^{\mrair}= p^{\mrair}_\mrbub-p^{\mrair}_\mratm =p_\mrhyd+p_\mryl-(p^{\mrv}_\mrbub-p^{\mrv}_\mratm)
\label{eq:2}
\end{equation}
with $p^{\mrv}_\mrbub$ and $p^{\mrv}_\mratm$ the vapor pressures in the bubble and surrounding atmosphere, respectively. Thus, the growth curve for water in Fig. \ref{fig:PQtot} should be shifted with respect to that for glycerol by the difference in vapor pressures $\Delta{p}^\mrv$. At small $Q_\mrtot$, we read off a pressure difference of circa $ 4600~\mathrm{Pa}$, which is indeed in agreement with $\Delta{p}^\mrv$: Experiments were performed at $~20^\circ$C room temperature and 20\% humidity, while in the bubble we expect to have 100\% humidity as well as an elevated temperature due to the focused halogen lighting. Assuming a temperature of $~33^\circ$C inside the bubble, the vapor pressure difference $\Delta{p}^\mrv$ takes the value observed in \figref{PQtot}.


These considerations allow for a prediction of the required hydrodynamic pressure for stabilization of a bubble, given the environmental conditions. Note that degassing of the water makes hardly any difference, because the relatively small volume of water in the microchannel is surrounded by gas-permeable PDMS, so that it quickly re-equilibrates (saturates with air). This can be estimated by the time it takes for gas to diffuse through the liquid layer of depth D (and thus establish quasi-static equilibrium) as $\tau_\mathrm{D}\sim \mathrm{D}^2/D_{\mathrm{air}}^{\mathrm{w}}\approx 5$\,s. Therefore we assume that water within the PDMS device gets quickly saturated with air, regardless of its initial state; the detailed time scale depends on the size of the PDMS layer and on the room humidity. These arguments lead us to the conclusion that the rate-limiting step in the growth process is the transport of air from the bubble through the PDMS layer.

\textcolor{black}{The air exchange at the rate $Q_\mrtot$ is driven by $\Delta{p}^\mrair$, the difference between the air partial pressure inside the bubble $p_{\mrbub}^{\mrair}$ and the atmospheric air partial pressure $p_\mratm^\mrair$; this exchange is slow and should follow a Darcy-like law\cite{darcy}, such that: 
\begin{equation}
Q_\mrtot={\Delta{p}^\mrair}/R_{\mrtot},
\label{eq:resistance}
\end{equation}
where $R_\mrtot$ represents the resistance encountered by the air as it is transported through the PMDS and the liquid layer. The value of $R_\mrtot$  depends on the geometry of the system and on a non-trivial combination of the permeabilities of PDMS and liquid to air. In the particular case of glycerol, given its negligible vapor pressure\cite{lide2004}, we can safely assume that $p_\mrbub^{\mathrm{v}}=0$ (see \eqref{2}).}

Beyond predicting stable bubbles, it is also valuable to describe the time scales of gas exchange for growing and shrinking bubbles, i.e., to quantify the resistance $R_\mathrm{tot}$ in \eqref{resistance} as a function of the bubble size and shape, which is itself time-dependent. The total gas flow rate $Q_\mathrm{tot}$ is the sum of the flow rate of gas leaving the bubble directly through the PDMS, $Q_\mrs$, and that of gas transported through the channel liquid (and from there through further PDMS layers surrounding the channel), $Q_\mrl$. The two contributions are governed by different geometries: the contact surfaces of gas with PDMS consists of the side channel surface and, if the bubble has grown beyond the side channel, the four contact areas with the main channel walls. By contrast, the gas/liquid interface is either flat or semicylindrical -- i.e., most of the solid angle of gas exchange is effected through the solid. Additionally, the fact that the size of the bubble is at least one order of magnitude smaller than the thickness of the surrounding PDMS walls, together with the good agreement of the bubble volume change with the radial-symmetry law $V\propto{t}^{3/2}$ (see \textcolor{black}{log-log plot in the} insert of \figref{TVgly}a and supplementary data), suggests an effective description of the bubble as an spherical gas object with an effective radius $r_\mrs$ (the position where the boundary conditions between bubble and PDMS are applied when the gas exchange is modeled as radially symmetric).

Assuming that the concentration profile of gas in the liquid is quasi-stationary in the spherical symmetry of this problem, closed solutions are available for the diffusion equation \cite{craft1959}. Combined with a boundary condition given by Henry's law at the gas-liquid interface, it is easy to derive the following formula for this portion of the transport:

\begin{equation}
R_\mrs = \Delta p^{\mrair}/Q_\mrs = \left(\Omega_\mrs K_\mrs r_\mrs\right)^{-1}
\label{eq:rs}
\end{equation}
where the constant $K_\mrs=\kappa/\mu$, with $\mu$ the viscosity of air and $\kappa$ the permeability of PDMS. Values for $K_\mrs$ are directly available in the literature, but vary according to the precise composition and cross-linking of the PDMS. We use the values of de Jong \emph{et al.} \cite{deJong2006} for the permeabilities for N$_2$ and O$_2$, weighted by their relative fractions, to obtain an effective value of 
$K_\mrs\approx 350$\,barrer $= 2.6\times 10^{-15}$\,m$^2$/(Pa\,s). The radial scale $r_\mrs$ is determined by the surface area $A_\mrs=\Omega_\mrs r_\mrs^2$ of the bubble-PDMS interface, if this transport affects a solid angle $\Omega_\mrs$. 

The transport of gas through the liquid is governed by the flux through the bubble surface. Focusing on the case where the bubble is semi-cylindrical of radius $a$, and assuming that the concentration profile of gas in the liquid is quasi-stationary in the cylindrical symmetry of this problem (cf.~\cite{carslaw1959}), it is easy to derive the following formula for this portion of the transport:
\begin{equation}
R_\mrl = \Delta p^{\mrair}/Q_\mrl = -\ln (a/r_{\mrint})\left(\pi K_\mrl \mathrm{D}\right)^{-1}\,.
\label{eq:rl}
\end{equation}

Here, $r_{\mrint}$ is the typical distance from the bubble center to the liquid/PDMS interface (so we set $r_{\mrint}=\mathrm{H}$), while the diffusive gas transport is governed by the constant $K_\mrl = v_\mathrm{m} D_{\mathrm{air}}^{\mathrm{w}} k_\mrH$, where $v_\mathrm{m}$ is the ideal gas molar volume, $D_{\mathrm{air}}^{\mathrm{w}}$ was given above and $k_\mrH$ is the Henry's law constant. For the system air/water $k_\mrH \approx 7.9 \times 10^{-4} \mathrm{~mol~m^{-3} \mathrm{Pa}^{-1}}$ and for air/glycerol $k_\mrH \approx 8.5 \times 10^{-4} \mathrm{~mol~m^{-3} Pa^{-1}}$. For the latter we estimated a value of $K_\mathrm{l}$ at least one order of magnitude smaller than that of air/water based on the values found in the literature \cite{rischbieter1996,thomas1965}  (see also the calculations in the supplementary material).


\begin{figure}[t]
 \includegraphics[width=\columnwidth]{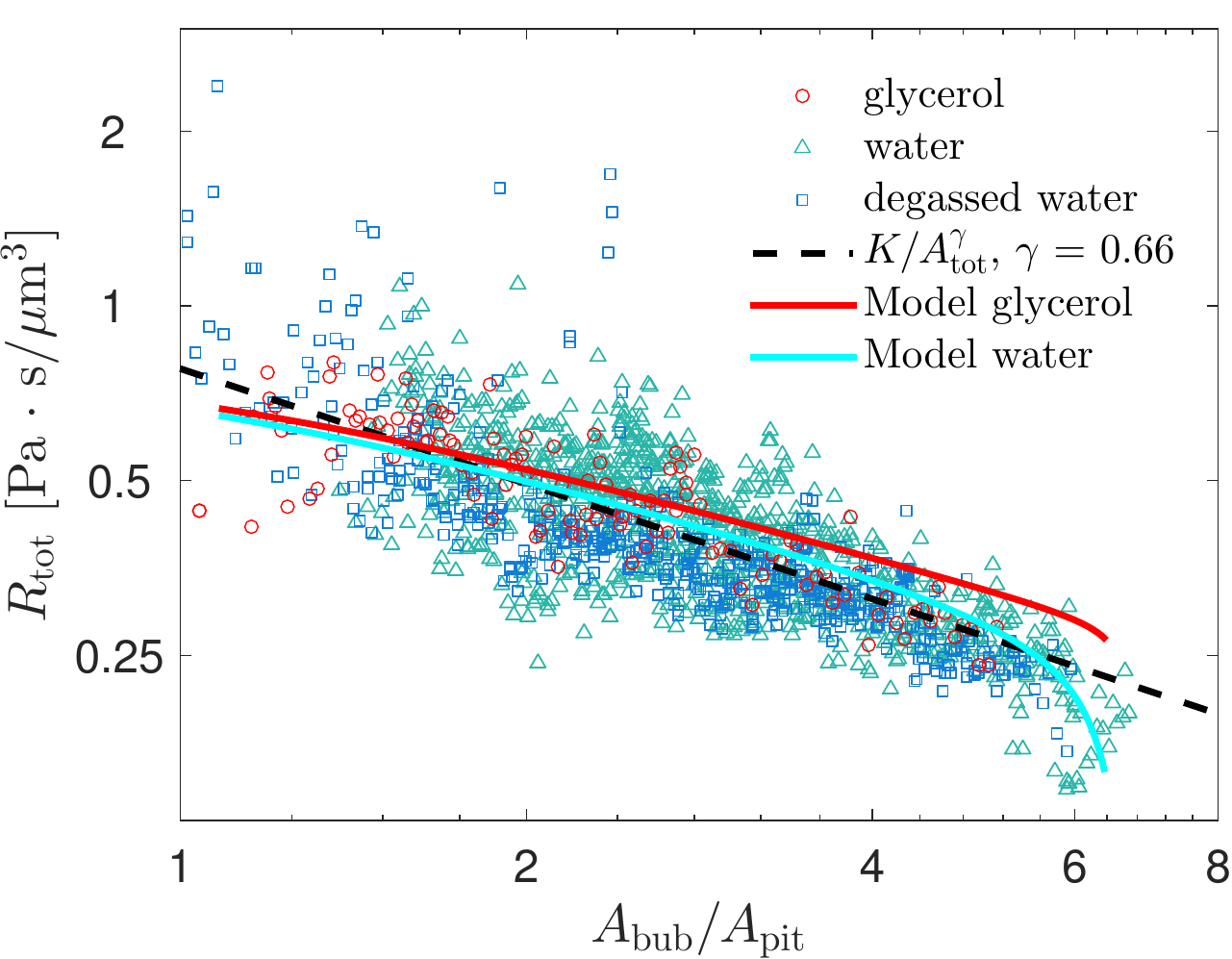}
\caption{\figlab{RA_all} Resistance to gas flow $R_\mathrm{tot}=\Delta{p}^\mrair/Q_\mrtot$ as a function of the bubble area in the experiments with glycerol, air-saturated water and degassed water (symbols).  The solid lines are theoretical results from the model developed in the main text. The dashed line is a power-law fit to the data corresponding to the equation: 
$R_\mathrm{tot}=K/A_\mrtot^{\gamma}$, with $\gamma=0.66$ and $K=1.17\cdot10^{13}~\mathrm{Pa\cdot s/m^{3-2\gamma}}$. \textcolor{black}{The data points correspond to all measured stages in the growing process for all given hydrostatic pressures (see \figref{TVgly}a).}}
\end{figure}

Combining  \eqref{rs} and \eqref{rl} we obtain a prediction for the dependence of the total transport resistance $R_{\mrtot}=\Delta p^\mrair/Q_{\mrtot}$ on the bubble radius $a$. Assuming that $\Omega_s=2\pi$ (partitioning the transport into half-spaces of solid and liquid transport), we can rewrite this relation in terms of $A_{\mrtot}=A_\mrl+A_\mrs$, where
\begin{equation}
A_\mrl=\pi a \mathrm{D}
\label{eq:al}
\end{equation}
and
\begin{equation}
A_\mrs=2\pi r_\mrs^2 = \pi a^2+2a\mathrm{D} +2(\mathrm{D}+\mathrm{W})\mathrm{L}\,.
\label{eq:as}
\end{equation}
A sketch with the area calculation is included in the supplementary material. Figure \ref{fig:RA_all} shows the experimental results for this relation $R_{\mrtot}(A_{\mrbub})$, using the permeability data given above for water, and merely correcting the data for glycerol by the vapor pressure difference. Also shown are the theoretical predictions from \eqref{resistance} to \eqref{as}, with $R_{\mrtot}^{-1} = R_{\mrs}^{-1} + R_{\mrl}^{-1}$, as well as the best-fit power law. As resistance is by definition independent of the materials involved, the agreement of values for glycerol and water are a successful test of consistency for both experiment and theory. Theory and experiment agree that this transport relation is somewhat steeper than what would be expected for a purely radial Darcy transport (which would result in  $\Delta{p}^\mrair/Q_\mrtot\propto 1/A_\mrtot^{1/2}$), which is attributable to the transport through the liquid. Without adjustable parameters, the theory developed above captures the magnitude of gas flow with good accuracy, including the strong drop in resistance around $A_{\mrbub}/A_{\mrpit}\approx 6$, where (in both experiment and theory) the growing bubble radius becomes comparable to $\mathrm{H}$, so that the vanishing liquid layer offers almost no resistance to gas transfer and the bubble grows faster. 

\begin{figure}[b]
 \includegraphics[width=\columnwidth]{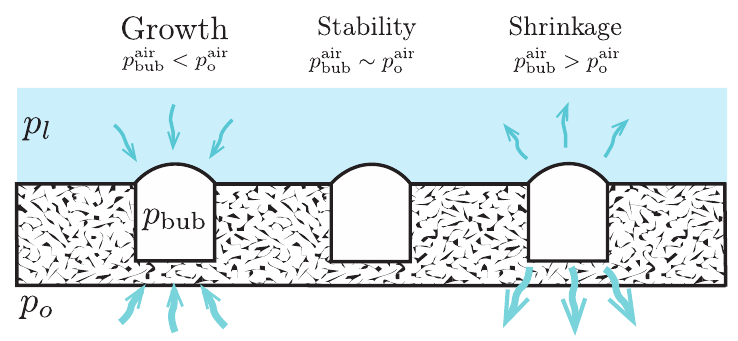}
\caption{\figlab{sketch} Illustration on the bubble stabilization mechanism based on the control of hydrostatic pressure of the liquid $p_\mrl$. The air flow through the solid/liquid media surrounding the bubble is driven by the difference between the air partial pressure inside the bubble $p^\mrair_\mrbub$ and the atmospheric air partial pressure $p^\mrair_\mro$.}
\end{figure}

\section{Conclusions}
\seclab{conclusions}
In cases where the substrate is totally impermeable to gas transfer (e.g.\ in silicon devices), the control of the bubble shape can be performed by either controlling the temperature or the amount of gas dissolved in the liquid, which is exchanged with the bubble by diffusion. In our experiments, we demonstrate that bubble growth and shrinkage in permeable materials is dominated by the diffusive transport of gas through its surrounding media, which actually allows for a higher degree of control and more accurate predictions when all transport processes are properly taken into account. By simply controlling the liquid's hydrostatic pressure, one can control the bubble's air partial pressure and its growth/shrinkage (see Fig. \ref{fig:sketch}). The critical value of the hydrostatic pressure for stability depends only on the Young-Laplace pressure of the bubble and on the vapor pressure of the liquid employed. The latter is a crucial, and unappreciated, element in this situation -- changing vapor pressure by moderate changes in temperature can have decisive effects on bubble stability (far surpassing the effect of temperature on gas solubility or other transport parameters). Likewise, relative humidity in the laboratory can be an important factor, so that reproducible experiments using microbubbles in PDMS devices with aqueous solutions should be performed at controlled humidity.

Note that in cases in which a net flow is driven inside the channel, a pressure gradient will develop along the channel and the pressure in the liquid will not be constant. Alternatively, the liquid pressure can then be controlled by connecting the outlet to a pressurized reservoir and adjusting the pressure inside the reservoir. Another option when using volatile liquids is to control the temperature of the system, and therefore to adapt the gas pressure inside the bubble mainly by altering the equilibrium vapor pressure. In the commonly used geometries in microfluidic channel devices, we find that the resistance to gas flow through the system can be accurately modeled if the bubble geometry is properly taken into account. \textcolor{black}{Additionally, the presence of a net flow might increase the relative importance of convective transport of air in comparison to the diffusive transport through the water liquid phase. However, since the key of the bubble stability relies on the transport through the PDMS rather than through the liquid phase, the influence of convective transport terms is minor (see calculations included in the supplementary material).}

In summary, in this paper we demonstrate how crucial is the choice of materials and liquids when working with microbubbles in microfluidic applications. We also show that, if the gas transfer problem is well understood, the permeability of commonly-used materials such as PDMS can actually become an advantage when working with multiphase flows in microsystems, membraneless exchangers, transport over bubble mattresses \cite{karatay2012} or bioreactors \cite{bioreactor}.

\section*{Acknowledgements}
Abundant discussions and technical help from Kilian Eckstein, Rune Barnkob, and Rocio Bola\~nos-Jim\'enez is greatly appreciated. A.V., M.R., A.M. and C.K. acknowledge financial support by the German Research Foundation grant KA 1808/17-1. S.H. acknowledges support by the National Science Foundation under grant number CBET- 1236141. The authors are grateful to Cheng Wang for providing the PDMS devices.


\providecommand*{\mcitethebibliography}{\thebibliography}
\csname @ifundefined\endcsname{endmcitethebibliography}
{\let\endmcitethebibliography\endthebibliography}{}

\end{document}